\documentclass[reprint,superscriptaddress,showpacs,amsmath,amssymb,aps,prl]{revtex4-1}
\usepackage{graphicx}
\usepackage{dcolumn}
\usepackage{bm}
\usepackage{hyperref}
\usepackage{subfigure}
\hypersetup{colorlinks=true, citecolor=blue, urlcolor=blue, linkcolor=blue}
\begin{document}
\title{Swarming transitions in hierarchical societies}

\author{Tingting Xue}\thanks{Equal contribution}
\affiliation{School of Physics and Information Technology, Shaanxi Normal University,
Xi'an 710061, P. R. China}
\author{Xu Li}\thanks{Equal contribution}
\affiliation{School of Physics and Information Technology, Shaanxi Normal University,
Xi'an 710061, P. R. China}
\author{Peter Grassberger}
\affiliation{JSC, FZ J\"ulich, D-52425 J\"ulich, Germany}
\author{Li Chen}
\email[Email address: ]{chenl@snnu.edu.cn}
\affiliation{School of Physics and Information Technology, Shaanxi Normal University,
Xi'an 710061, P. R. China}

\begin{abstract}
Social hierarchy is central to decision-making in the coordinated movement of many swarming species. 
Here we propose a hierarchical swarm model in the spirit of the Vicsek model of 
self-propelled particles. We show that, as the hierarchy becomes important, the swarming transition changes from the weak first-order transition observed for egalitarian populations, to a stronger first-order transition for intermediately strong hierarchies, and finally the discontinuity reduces till vanish, where the order-disorder transition appears to be absent in the extremely despotic societies. Associated to this we observe that the spatial structure of the swarm, as measured by the correlation between the density and velocity fields, is strongly mediated by the hierarchy.  A two-group model and vectorial noise are also studied for verification. Our results point out the particular relevance of the hierarchical structures to swarming transitions when doing specific case studies.
\end{abstract}

\date{\today }
\maketitle

\emph{Introduction} ---
Collective motion is one of most spectacular and fascinating emergent
behaviors in nature, as exhibited in insects, bird flocks, fish shoals, and herds of ungulates, among 
others~\cite{vicsek2012collective}. While detailed case studies are preferred in general by 
biologists~\cite{krause2002living,sumpter2010collective,buhl2006disorder}, physicists usually seek for minimal models with the hope 
that there are universal features behind seemingly diverse observations, and simple models are sufficient 
to capture the fundamental laws~\cite{marchetti2013hydrodynamics,chen2017weak}. A prototype model of the second kind, 
which considerably advanced our understanding of collective motion, is the Viscek model~\cite{vicsek1995novel} 
inspired by statistical physics, and its many variants~\cite{chate2008modeling}. A major concern here is 
the nature of swarming transitions between the ordered and disordered states of movement. The second-order 
phase transition (PT) claimed in the original work was later challenged by 
Ch\'ate \emph{et.al}~\cite{gregoire2004onset,chate2008collective}, by showing that the observed continuous 
nature is actually due to finite-size effects, and a first-order transition should be expected in the 
thermodynamical limit if only with local interactions -- in line with some theoretical 
studies~\cite{bertin2006boltzmann,bertin2009hydrodynamic,ihle2011kinetic}. Note that, analogous to the 
identical particle assumption in statistical mechanics, individuals in these models are supposed to be 
indistinguishable and thus equally important in decision-making for the movement coordination.

While the collective movements for some species can indeed be described as an equally shared 
consensus~\cite{strandburg2015shared, conradt2007democracy}, many more are based on partially shared or 
even unshared consensus decision-making~\cite{sueur2008shared, conradt2005consensus}, where only a tiny 
fraction of individuals (or even a single one) lead the group movement. This is particularly true  when 
hierarchical social structures are present. For example, recent experiments with high-resolution GPS data 
have revealed well-defined hierarchical structures among homing pigeons~\cite{nagy2010hierarchical} and 
migratory white storks~\cite{flack2018local}, where a small number of leaders direct the group flight. 
Highly asymmetrical dominance is also revealed in mammals such as African elephants~\cite{mccomb2001matriarchs}, 
gray wolves~\cite{peterson2002leadership}, and primates~\cite{schaller1963mountain, king2008dominance}, 
as well as in fish schools~\cite{jiang2017identifying} and honeybee swarms~\cite{seeley2014honeybee} \emph{etc}. 
However, a generic model that interpolates from egalitarian to despotic swarms is still lacking. Revealing 
how the hierarchy impacts on the collective motion remains a crucial challenge for understanding hierarchical 
societies in general.

In this work, we fill this gap by introducing a hierarchical swarm model, called the hierarchical Viscek model (HVM), and investigate the impact of the hierarchy on the nature of order-disorder transitions. By tuning the swarm from egalitarian to despotism, we show numerically that the swarming transition changes non-monotonically from a weakly first-order PT to stronger ones for intermediate levels of hierarchical impact, and finally the discontinuity shrinks and disappears. Detailed analysis of the microscopic structures shows that the altered nature is due to hierarchy-induced changes in correlations between the density and orientation fields. 

\begin{figure}[htbp]
\centering
{\label{Fig.1a}\includegraphics[width=4.2cm]{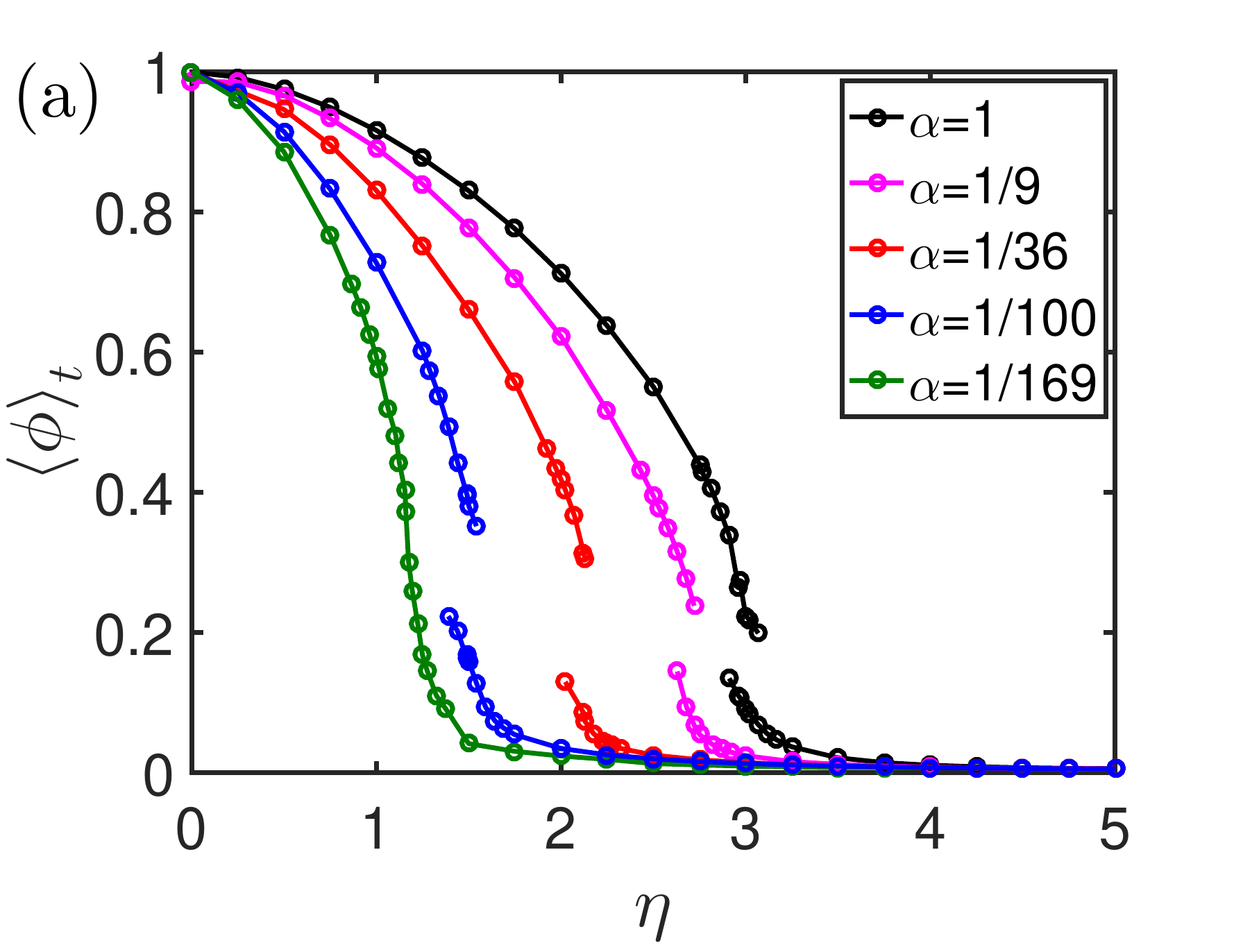}}
{\label{Fig.1b}\includegraphics[width=4.2cm]{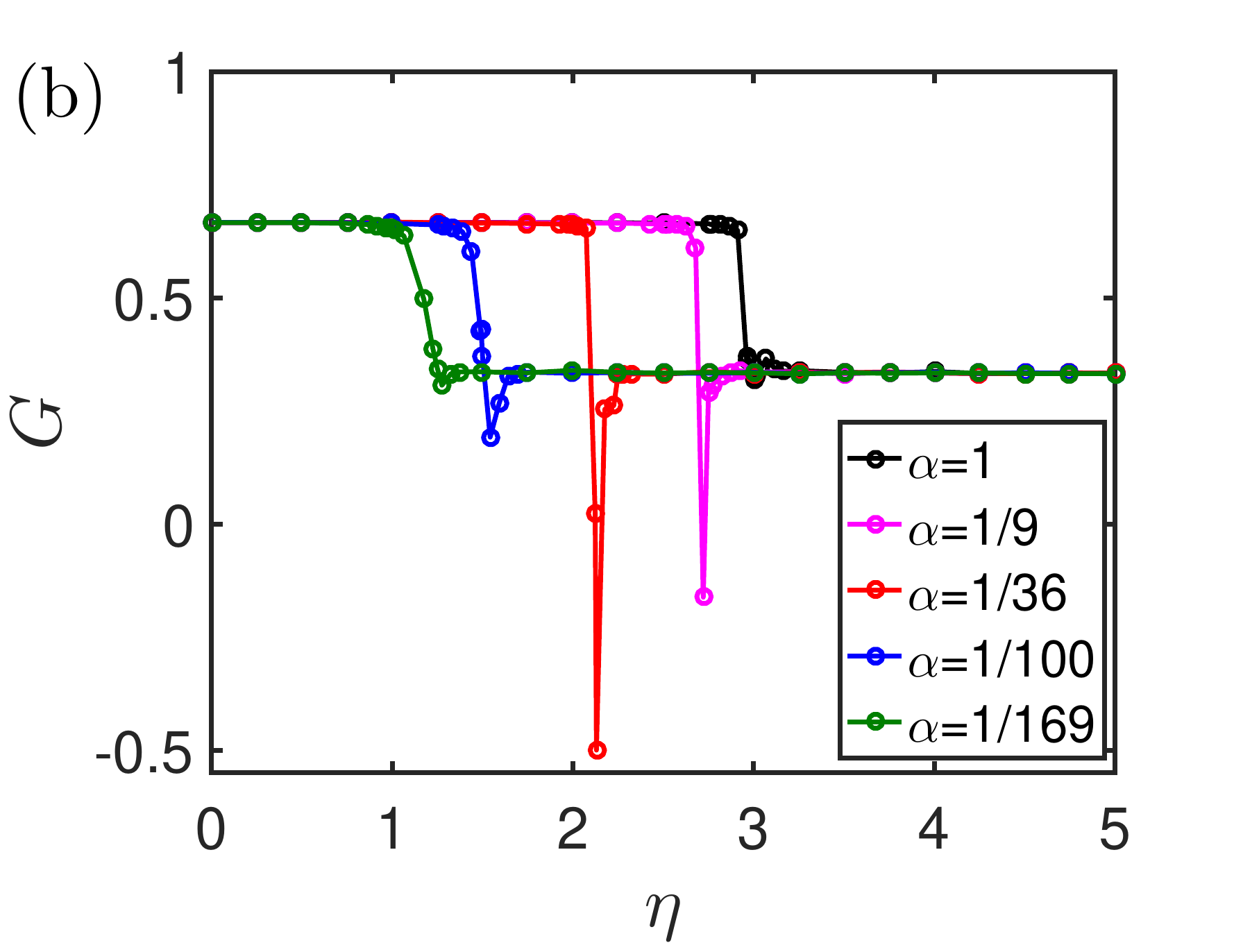}}
{\label{Fig.1c}\includegraphics[width=4.2cm]{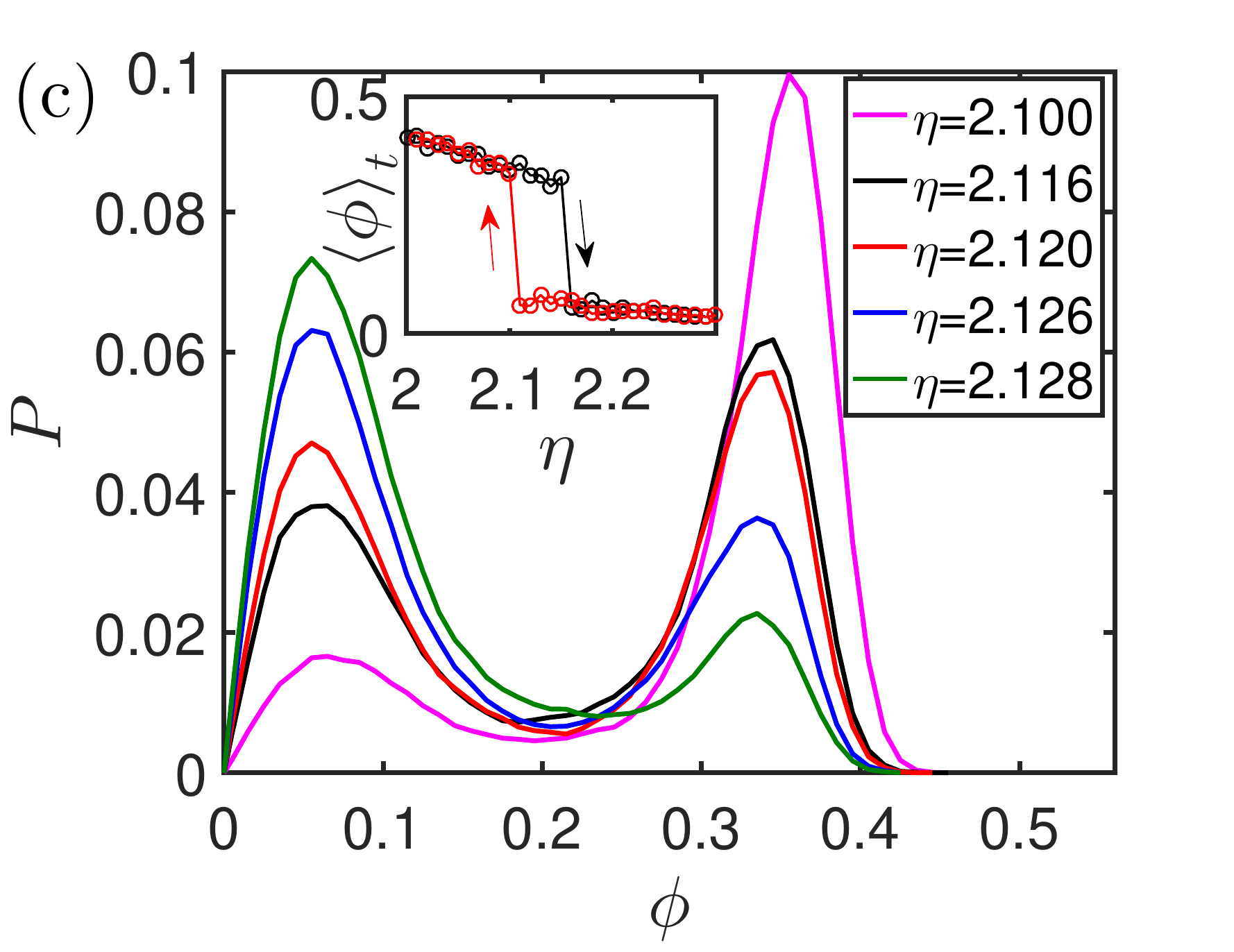}}
{\label{Fig.1d}\includegraphics[width=4.2cm]{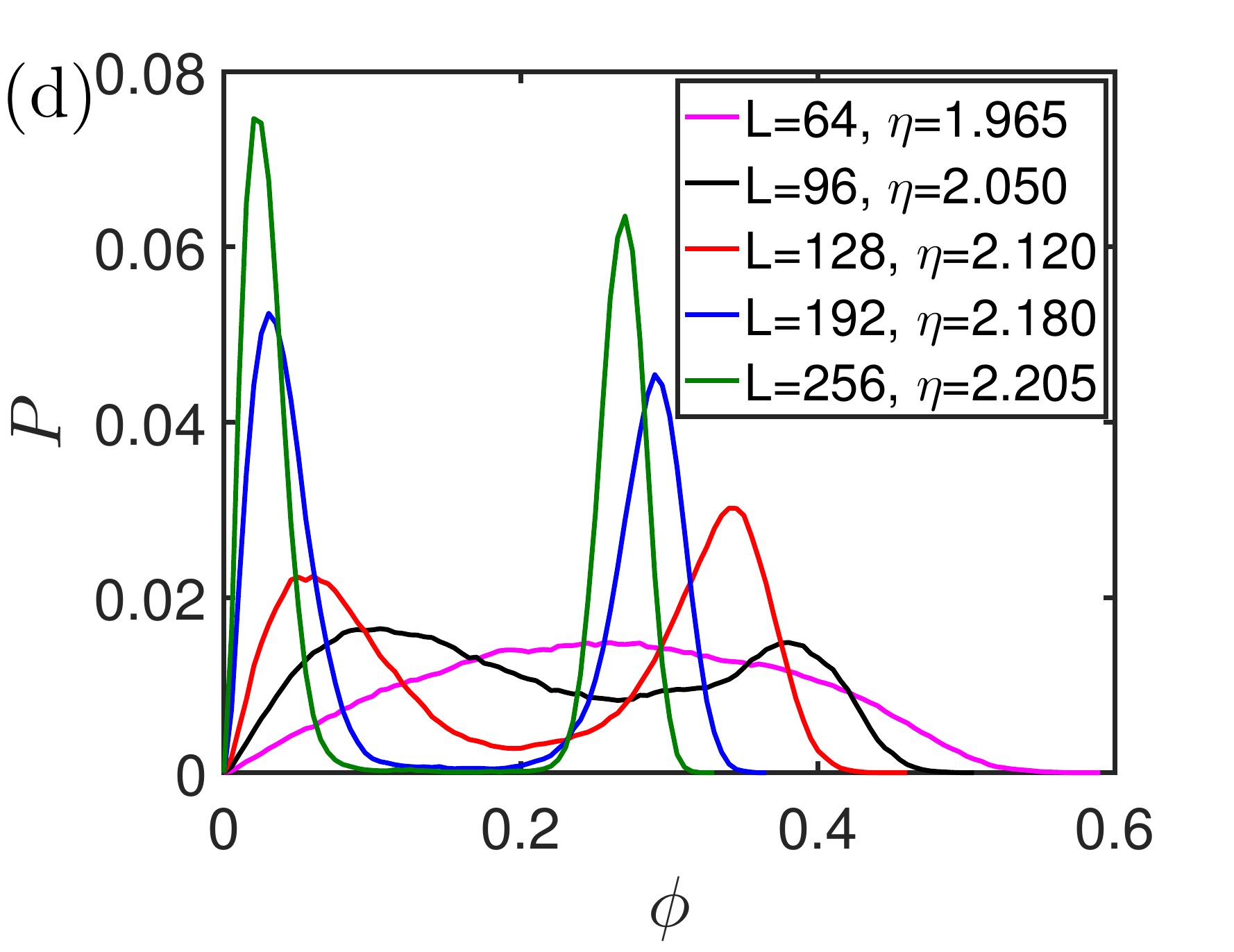}}
\caption{(Color online) Phase transitions of hierarchical Vicsek model (HVM). (a) The average OP $\left\langle\phi\right\rangle_t$ versus the noise amplitude $\eta$. ($\left\langle\phi\right\rangle_t$ is computed as following: if the PDF of $\phi$ is single-peaked, then we average all data; but if the profile is bimodal, as cases shown in (c), we divide the data at the valley point, and do average respectively for the ordered and disordered state, two points are then obtained for each $\eta$.) (b) Binder cumulant $G$ versus $\eta$, with same symbols and colors as in (a). Time average over $1\times10^{6}$ time steps. (c) PDF of $\phi$ near the  transition  point with $\alpha=1/36$. The hysteresis in the inset is shown with a ramp rate of $1/3\times10^{-6}$ per time step (each data average over $3\times10^{5}$ time steps). 
(d) The finite size effect near the transition point with $\alpha=1/36$. 
Parameters: $\rho=2$, $v_{0}=0.5$, $L=128$ for (a)-(c).}
\label{Fig.1}
\end{figure}

\emph{The hierarchical Viscek model} --- As in the standard Vicsek model (SVM)~\cite{vicsek1995novel}, 
$N$ pointwise particles labeled as $1,2,...,N$ are randomly placed on a two-dimensional domain with size 
$L\times L$ with periodic boundary conditions. They move synchronously at discrete time steps by a fixed 
distance $v_0\Delta t$, where $v_0$ is the velocity defined as the length of displacement per time step $\Delta t$, fixed at unity. Each particle $i$ is endowed with an angle $\theta_{i}$ that determines the direction of the movement during the next time step, and its update is determined by the orientations of its neighbors 
(defined as particles within a unit circle centered around particle $i$, including itself). In the SVM, the influence of the neighbors 
is through an average angle 
\begin{equation}
	\langle\theta_{i}(t)\rangle_r=\Theta[\sum_{j:\;d_{ij}<1} \mathbf{v}_j(t)] \quad {\rm (SVM)},    
	\label{eq:SVM-theta}
\end{equation}
where $\Theta [\mathbf{v}]$ represents the angle of vector $\mathbf{v}$ and $d_{ij}$ is the distance between 
particle $i$ and $j$. This is the main place where the HVM deviates from the SVM. Instead of the simple 
sum over all neighbors, we now order all particles by their hierarchical rank, with $j=1$ being highest and 
$j=N$ the lowest. While neighbors with rank $j<i$ have full influence on particle $i$, the influence of a 
lower-ranked neighbors is reduced by a factor $\alpha <1$. Instead of Eq.~(\ref{eq:SVM-theta}) we have thus
\begin{equation}
        \langle\theta_{i}(t)\rangle_r=\Theta[\!\!\!\sum_{d_{ij}<1,\;j\leq i} \!\!\!\mathbf{v}_j(t) + \alpha\!\!\! 
	                                      \sum_{d_{ij}<1,\;j>i} \!\!\!\mathbf{v}_j(t)] \quad {\rm (HVM)}.
        \label{eq:HVM-theta}
\end{equation}

The evolution is then the same as in the SVM,
\begin{equation}
\mathbf{x}_{i}(t+\Delta t)=\mathbf{x}_{i}(t)+\mathbf{v}_{i}(t+\Delta t)\Delta t,
\label{eq:SVM1}
\end{equation}
\begin{equation}
\theta_{i}(t+\Delta t)=\langle \theta_{i}(t)\rangle_r+\eta\xi_i(t).
\label{eq:SVM2}
\end{equation}
Here the key ingredient is the competition between the tendency towards local alignment and the angular noise $\xi_i(t)$ that might come from external perturbations and/or from uncertainties in individual's perception, chosen from a uniform distribution within the interval $[-1/2, 1/2]$, and $\eta$ is the noise amplitude.

In the limit $\alpha\!=\!1$ we recover the standard Viscek model, which is egalitarian by nature. Notice also that we 
could have implemented in our model more complicated hierarchical structures, by replacing simply the two values 
$(1, \alpha)$ by any function $\alpha(i,j)$.

In the absence of noise ($\eta\!=\!0$), all particles tend to align perfectly, while for the maximal noise ($\eta\!=\!2\pi$) 
they essentially make random walks. The transition between these two extremes can be conveniently measured by the 
order parameter (OP) defined as $\phi(t)\equiv\frac{1}{Nv_{0}}\left|\sum_i\mathbf{v}_{i}\right|$, and its temporal 
average $\left\langle\phi\right\rangle$. To monitor jumps of $\left\langle\phi\right\rangle$, a useful quantity 
is the so-called Binder cumulant $G(\eta,L)=1-\left\langle \phi^{4}\right\rangle/(3\left\langle \phi^{2}\right\rangle^{2})$,
which exhibits a characteristic minimum due to the phase coexistence in first-order phase transition and the minimum is expected to fall to negative values for strong discontinuities. In the 
ordered phase we expect roughly Gaussian distributions of $\phi(t)$ and thus $G\thickapprox2/3$, while in the 
disordered phase $G\thickapprox1/3$ in two dimensions.

\emph{Results and analysis} --- 
Varying the hierarchical coefficient $\alpha$, we observe a rich 
spectrum of phase transitions (PTs) as a function of noise intensity $\eta$, see Fig.~\ref{Fig.1}a. For the 
egalitarian case where $\alpha=1$, a first-order phase transition is seen but the gap between the two branches 
is small, as is also the decrease in the Binder cumulant $G$ (Fig.~\ref{Fig.1}b). Therefore it is weak, as also
found in previous studies \cite{gregoire2004onset,chate2008collective}. As $\alpha$ is decreased, the critical noise also $\eta_c$ decreases,
the gap becomes larger and the minimum of $G$ becomes deeper, suggesting an increasingly stronger PT. Around 
$\alpha\approx 1/36$, this enhancement becomes maximal. Decreasing $\alpha$ further, the gap shrinks
again and the minimum of $G$ becomes again shallower. Finally, for $\alpha \rightarrow 0$ the curve of 
$\left\langle\phi\right\rangle$ against $\eta$ approaches continuous and the minimum of $G$ is at $G\approx 1/3$. 
Note that, the transitions at $\alpha>0$ are still of discontinuous nature in the thermodynamic limit, but they are so weak that continuous PTs are expected in reasonably large swarms.
At $\alpha=0$, the strict continuity is observed, but there seems no order-disorder transition at finite $\eta_c$, therefore it's hard to conclude second order PTs in thermodynamic limit at this extreme. Detailed analysis is provided in Sec. I of SM~\cite{SM}.

The observation of enhanced discontinuities is strengthened by the bimodal probability distribution 
functions (PDFs) as shown in Fig.~\ref{Fig.1}c. 
The peak at smaller values of $\phi$ corresponds to disordered motion, the other to the ordered phase. As 
expected, the ordered phase shrinks when the noise is increased, while the disordered phase expands. 
Another hallmark of first-order PTs, also clearly seen in Fig.~\ref{Fig.1}c, is the presence of the hysteresis.
Fig.~\ref{Fig.1}d provides further evidence by the finite size effect analysis, showing more and more sharp bimodal peaks as the system becomes larger.

\begin{figure}[t]
{\label{Fig.2a}\includegraphics[width=4.2cm]{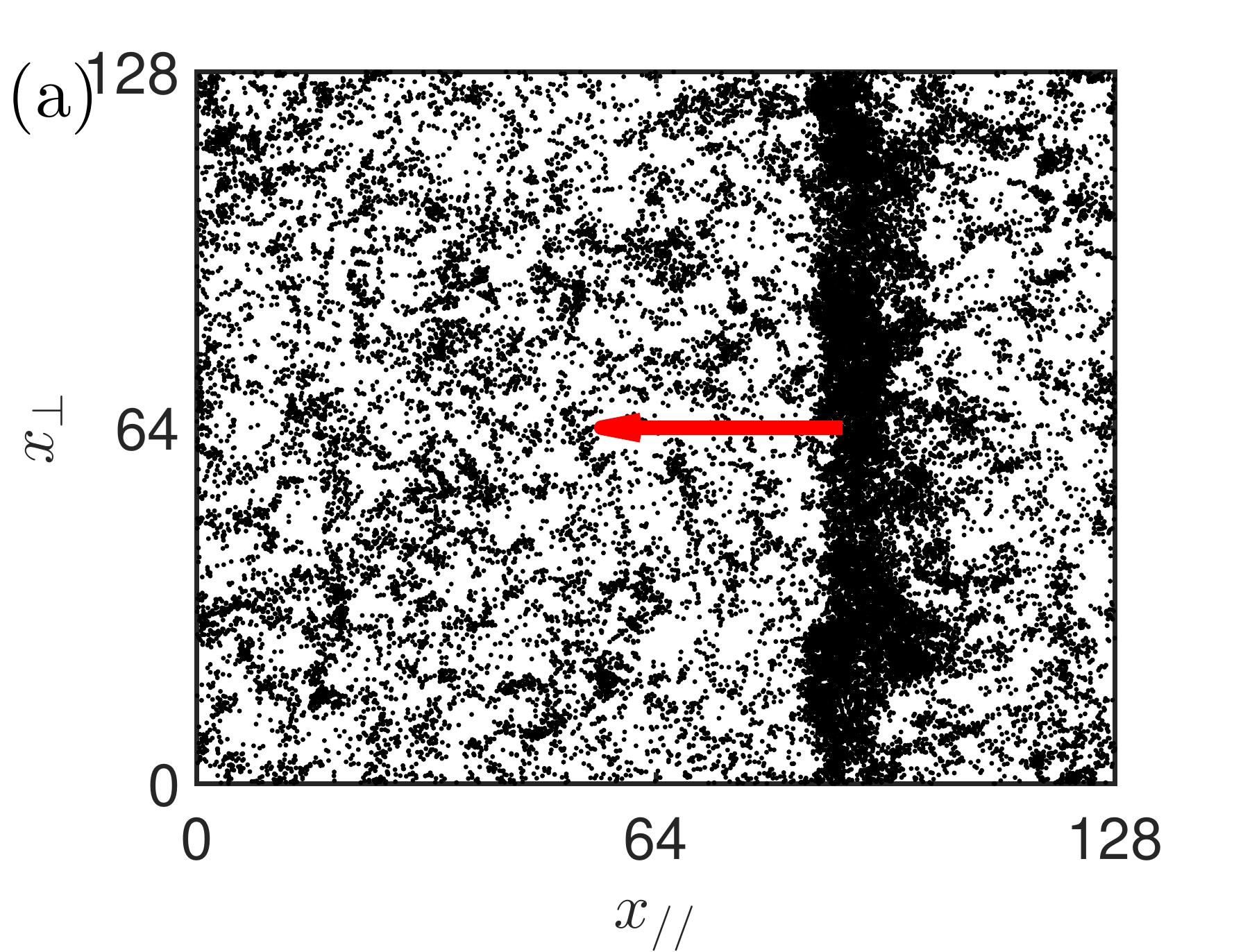}}
{\label{Fig.2b}\includegraphics[width=4.2cm]{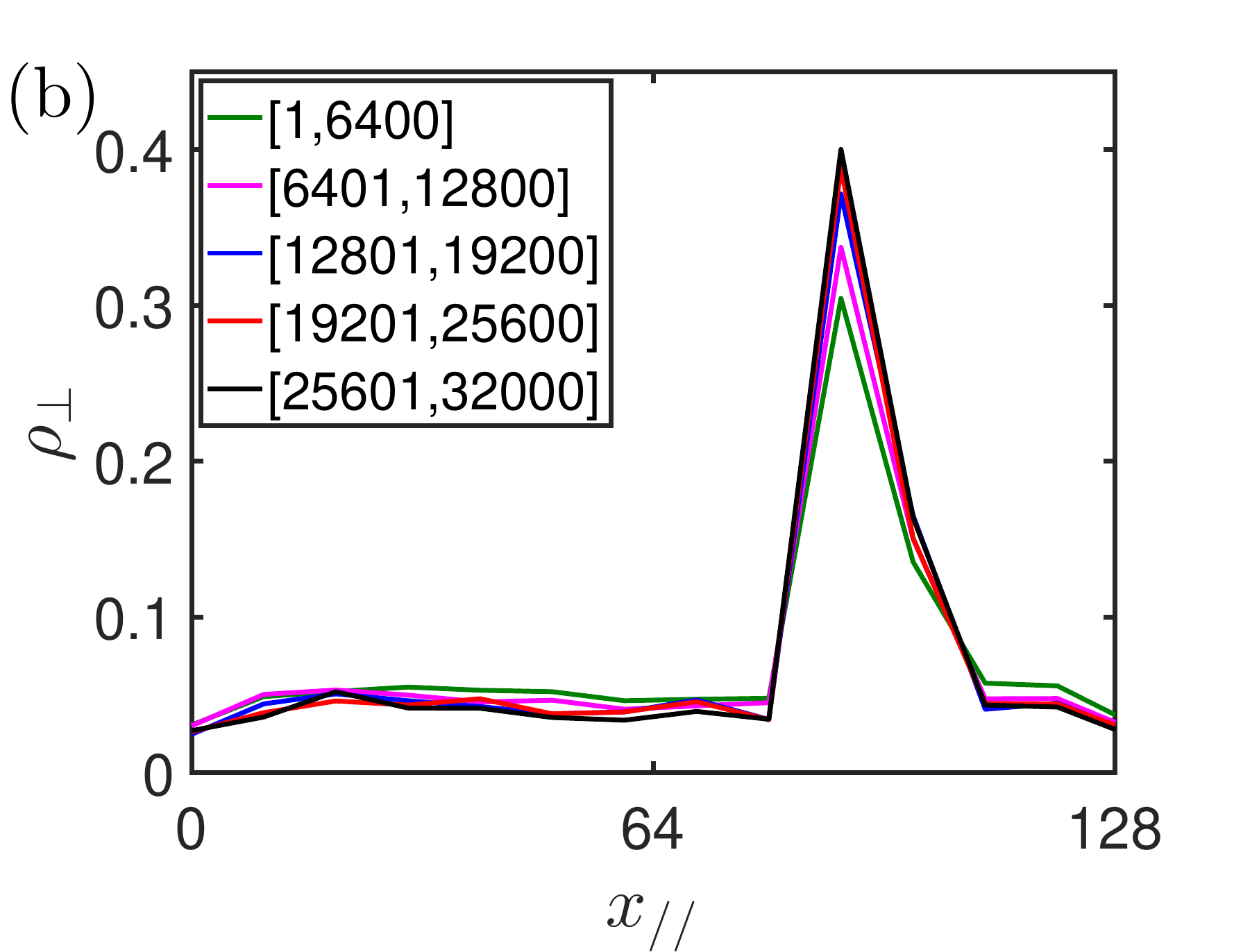}}
\caption{(Color online) Band structure in our HVM.
(a) A typical snapshot in the ordered state within the bistable region. Points represent the position of individual particles and the red arrow indicates the average direction of motion.
(b) Density profiles \textcolor{blue}{$\rho_{\perp}(x_{||},t)$} along the direction of $x_{||}$ for five subgroups, each contains 6400 particles.
Parameters: $\alpha=1/36$, $\eta=2.12$.}
\label{Fig.2}
\end{figure}

To understand how the hierarchy affects the swarming transitions, we first look at how the 
spatial distribution is influenced. A known feature of the SVM is that there exist localized, high-density, 
traveling bands corresponding to the ordered, symmetry-breaking phase. They are metastable on long time 
scales, i.e. they dissolve and reappear from time to time. 
As seen in Fig.~\ref{Fig.2}a, this is even more pronounced for the HVM at $\alpha=1/36$,
where the transition is most abrupt.
To see the impact of the hierarchy, we divide the population into five subgroups based on their labels, 
and compare their density profiles $\rho_{_\perp}$ along the average moving direction (Fig.~\ref{Fig.2}b). 
While a kink-like profile is observed for all groups, it is most 
pronounced for the low-rank group and least pronounced for the group with highest rank
(more details see Sec. II of SM~\cite{SM}). Intuitively, individuals on top of the hierarchy are least sensitive to their neighbors. Thus they feel the weakest collective force, and they have the least tendency to be trapped in the bands. For 
individuals of low rank, the opposite is true.

This explains immediately why the PT approaches continuous as $\alpha\rightarrow 0$
by examining the band stability. In this limit, high-ranking individuals completely 
ignore nearly all others and can only be influenced by neighbors of even higher rank. Assume 
now that there exists a band. The top ranking particle is blind to it, and will therefore soon leave it.
But then the second-top ranking particle becomes top-ranking in the band, and will also leave it.  As the
departure process repeats, the band will finally dissolve. Thus, bands become unstable as
$\alpha\rightarrow 0$, nor any other ordered structures could be expected.
A continuous PT is expected instead. This hierarchy-dependent departure is numerically confirmed 
in Sec. III of SM~\cite{SM}.

\begin{figure}[t]
\centering
{\label{Fig.3}\includegraphics[width=5.8cm]{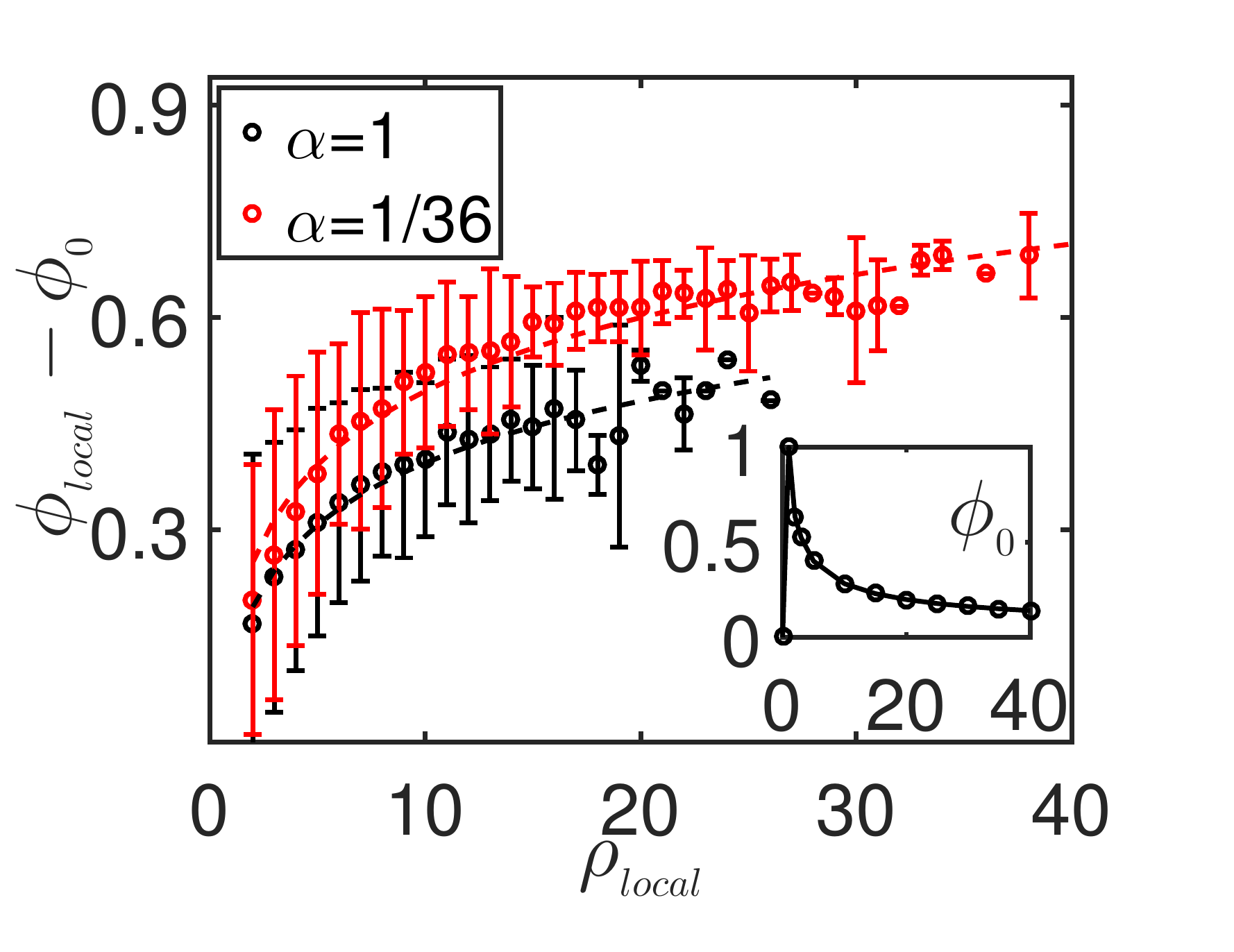}}
\caption{(Color online) Correlation between density and velocity fields.
Effective local OP $\phi_{_{local}}-\phi_{0}$ versus local density $\rho_{_{local}}$ (mean $\pm$ standard deviation). The correlation can be fitted by $\phi_{_{local}}-\phi_0 \sim\gamma*\lg\rho_{_{local}}$, where $\gamma\approx 0.29(1)$ for $\alpha=1$, and 0.34(6) for $\alpha=1/36$.
Inset shows the finite particle effect for the local OP $\phi_0$, which is computed by averaging an assemble of randomly orientated particles for different densities $\rho_{_{local}}$ in a cell.
Parameters: $\eta=2.97$ for $\alpha=1$ and $\eta=2.12$ for $\alpha=1/36$.}
\label{Fig.3}
\end{figure}

To understand the enhanced discontinuities, we turn to study the coupling between the particle density and orientation fields.
A consensus regarding the mechanism for band dynamics is that it is due to the intimate feedback within these two fields that in a cascaded manner leads to the band emergence and disappearance. Consider a moving patch with a slightly higher density than its surroundings near the transition points, since its orientation field is better aligned, it is more likely to attract particles come across, which in turn makes the patch more ordered and is again even more likely to attract particles. Band dissolution occurs in just opposite cascades. 
The feedback between density and orientation fields makes the bands form and dissolve recurrently. 
Fig.~\ref{Fig.3} shows that there is indeed a positive correlation between the local density $\rho_{_{local}}$ (defined as the particle number in a cell) and the effective local order parameter $\phi_{_{local}}-\phi_0$ ($\phi_{_{local}}$ is defined as the spatially averaged OP in cells of size $1\times1$, details see Sec. IV of~\cite{SM}). Notice that, $\phi_0$ is the background value of $\phi_{_{local}}$ that comes from finite particle effect --- a smaller number of particles always produce finite $\phi_{_{local}}$ even if their headings are completely uncorrelated (see the inset). Therefore $\phi_{_{local}}-\phi_0$ measures the degree of order that purely comes from the coordination process. By comparison, the case with hierarchical impact hold a significant improvement in the orientation field given the same density field. More importantly, there is a considerably high density region $\rho_{_{local}}\!>\!25$ that only appears in hierarchical swarms, and these highly dense patches usually are most crucial for nucleation processes. These features can also be validated by comparing the distributions of local order parameters, and even slightly enhance giant number fluctuations (see Sec. IV of SM~\cite{SM}).

Additional insight from a network perspective~\cite{aldana2003phase,aldana2007phase,miguel2018effects,huepe2011adaptive,chen2016adaptive} for the enhanced discontinuity is also given in Sec. V of SM~\cite{SM}, where swarming particles are taken as networked vectors $e^{i\theta}$. The discontinuity in swarming transition corresponds to the order parameter difference of two networks, one for disordered states, the other for band state. A plausible argument regarding the absence of order-disorder PT in the case of $\alpha=0$ is also provided.

\begin{figure}[t]
\centering
{\label{Fig.4a}\includegraphics[width=4.2cm]{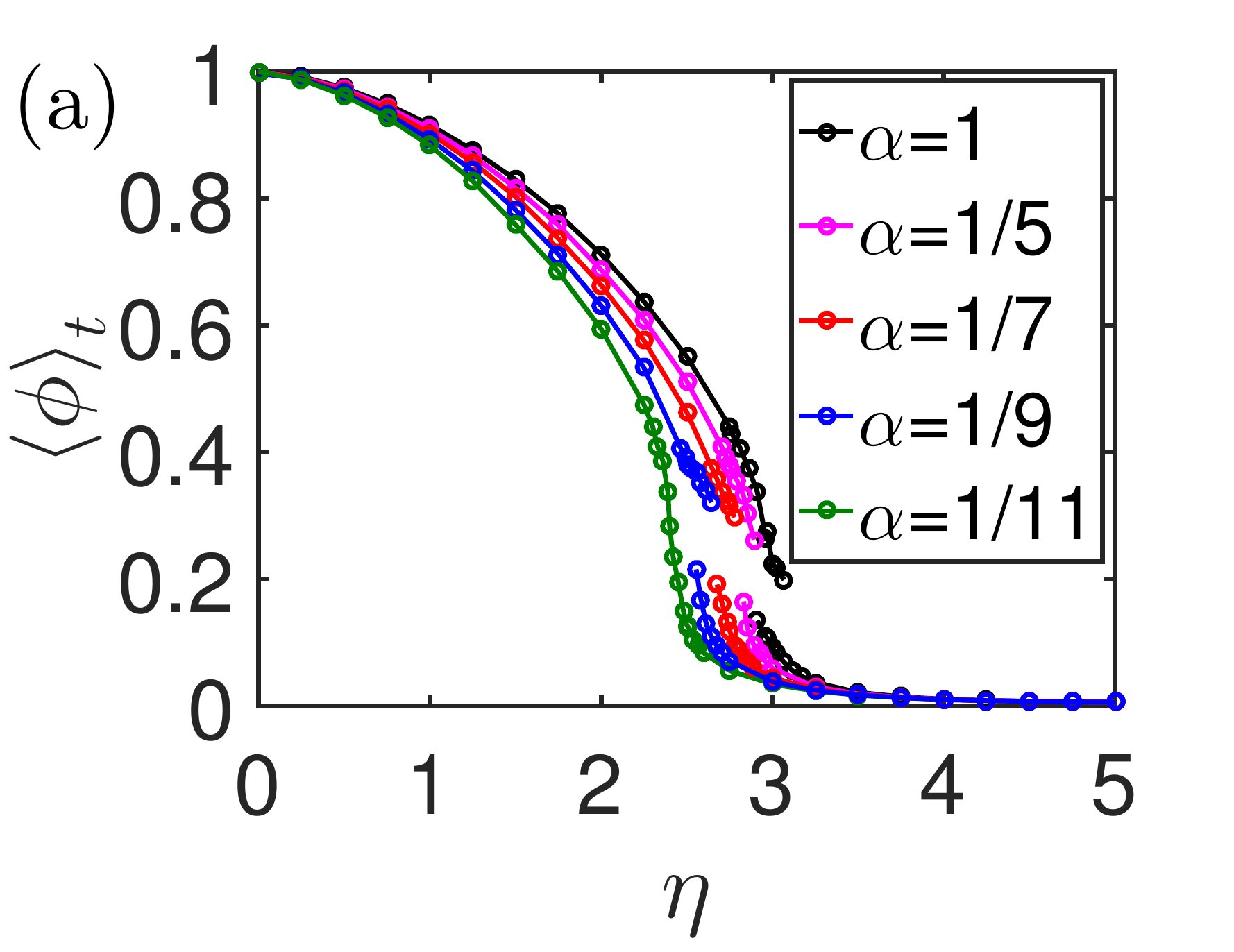}}
{\label{Fig.4b}\includegraphics[width=4.2cm]{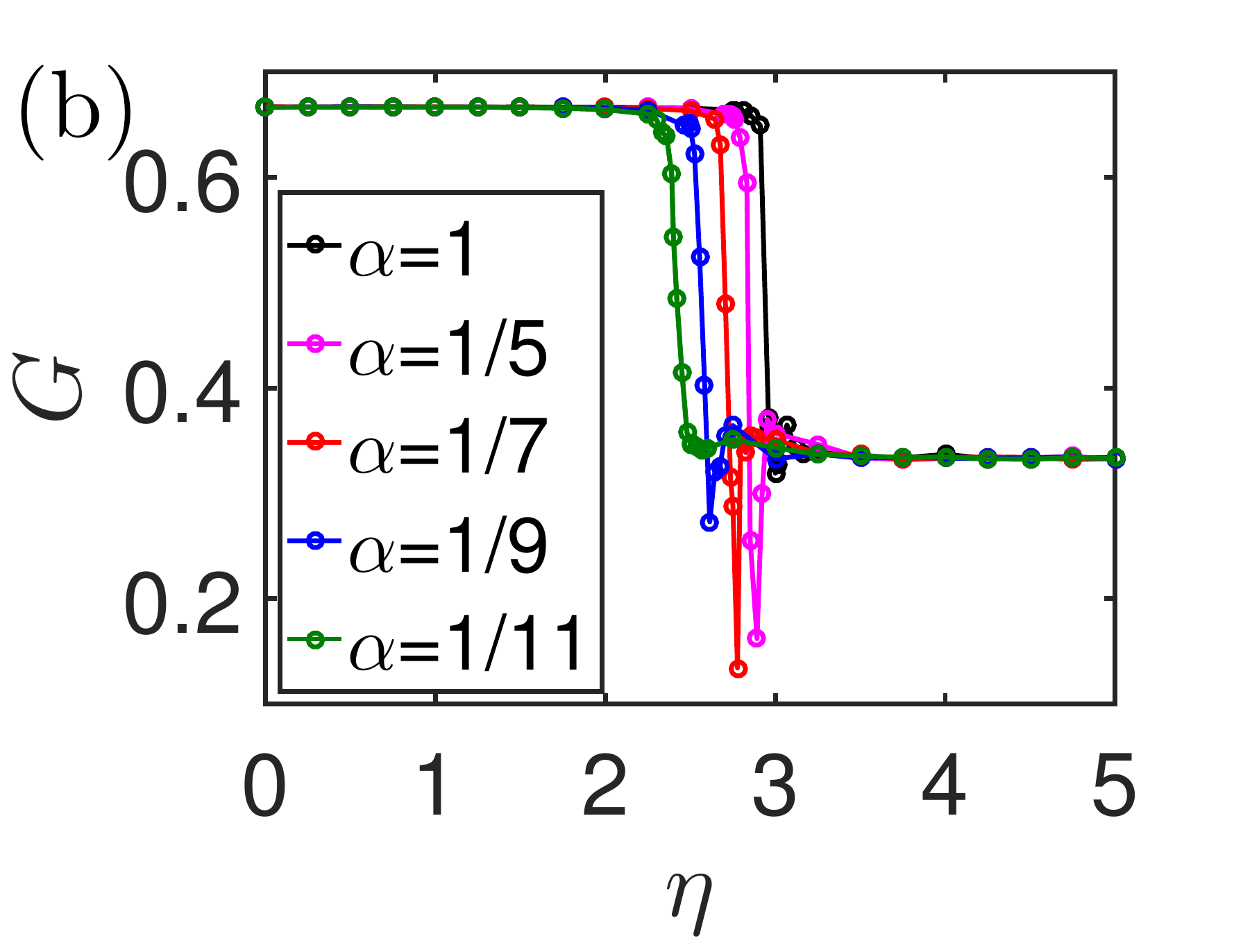}}
\caption{(Color online) Phase transitions of two-group model.
(a) $\left\langle\phi\right\rangle_t$ versus the noise strength $\eta$.
(b) Binder cumulant $G$ versus $\eta$.
Parameters: $h=0.1$, data averaged over $10^{6}$ time steps after transient.}
\label{Fig.4}
\end{figure}
\emph{Generalization 1. A two-group model} --- For generality, we study a further simplified HVM, where the population is only divided into two subgroups~\cite{baglietto2013gregarious,netzer2019heterogeneous}: one with high rank with fraction $h$, the other of low rank with $1-h$. The alignment is as
\begin{small}
\begin{equation}
\left\langle \theta_{i}(t)\right\rangle _{r}=\Theta[\alpha\!\!\!\!\sum\limits_{d_{ij}<r,j>M}\!\!\!\!\!\mathbf{v}_{j}(t)+\!\!\!\sum\limits_{d_{ij}<r,j\leq M}\!\!\!\!\!\mathbf{v}_{j}(t)],
\end{equation}
\end{small}
where $M\!=\!Nh$ is the size of the high-rank group. Fig. \ref{Fig.4}a shows a similar transition in the case of $h=0.1$. A weak 1st PT --- strong 1st PT --- a continuous-like PT scenario is seen when $\alpha$ decreases, in line with the observations of Binder cumulant $G$ in Fig. \ref{Fig.4}b. Though the enhancement in discontinuity is less significant than the case of HVM and the optimal value is also different, now around $1/7$.

\emph{Generalization 2. Hierarchical swarm model with vectorial noise} --- Besides the Vicsek model, Chat\'e et al. proposed another well-adopted swarm model \cite{gregoire2004onset,chate2008collective}, where the major difference is the way of noise implementation. Here, the evolution of the hierarchical swarm is exactly the same as HVM described by Eq.~(\ref{eq:HVM-theta}-\ref{eq:SVM2}) except the noise term is now replaced by the \emph{vectorial} type $\eta N_i\xi_i(t)$, where $N_i$ is the number of particles in $i$'s neighborhood. In HVM or SVM, the scalar noise can be taken as the action errors in following their locally averaged headings, which is also termed as \emph{angular} noise. The vectorial noise instead can be interpreted as the perception uncertainties in each pairwise interaction. While the intensity of angular noise is independent of the local order, the relative intensity of vectorial noise however decreases as the local alignment becomes strong. This property makes the vectorial noise more easier to trigger the band formation/deformation, leading to stronger first-order phase transitions than the angular noise case.

\begin{figure}[t]
\centering
{\includegraphics[width=4.2cm]{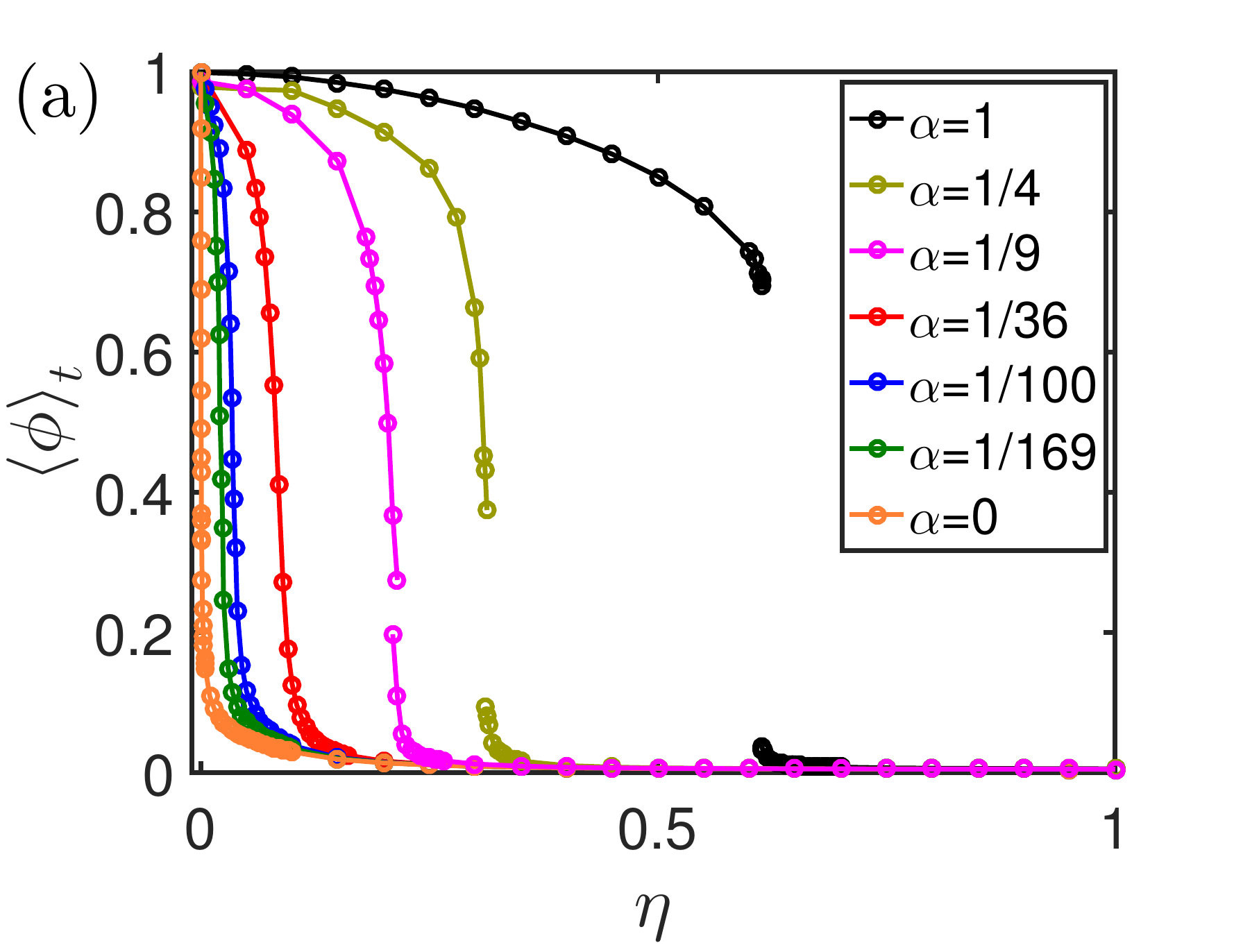}}
{\includegraphics[width=4.2cm]{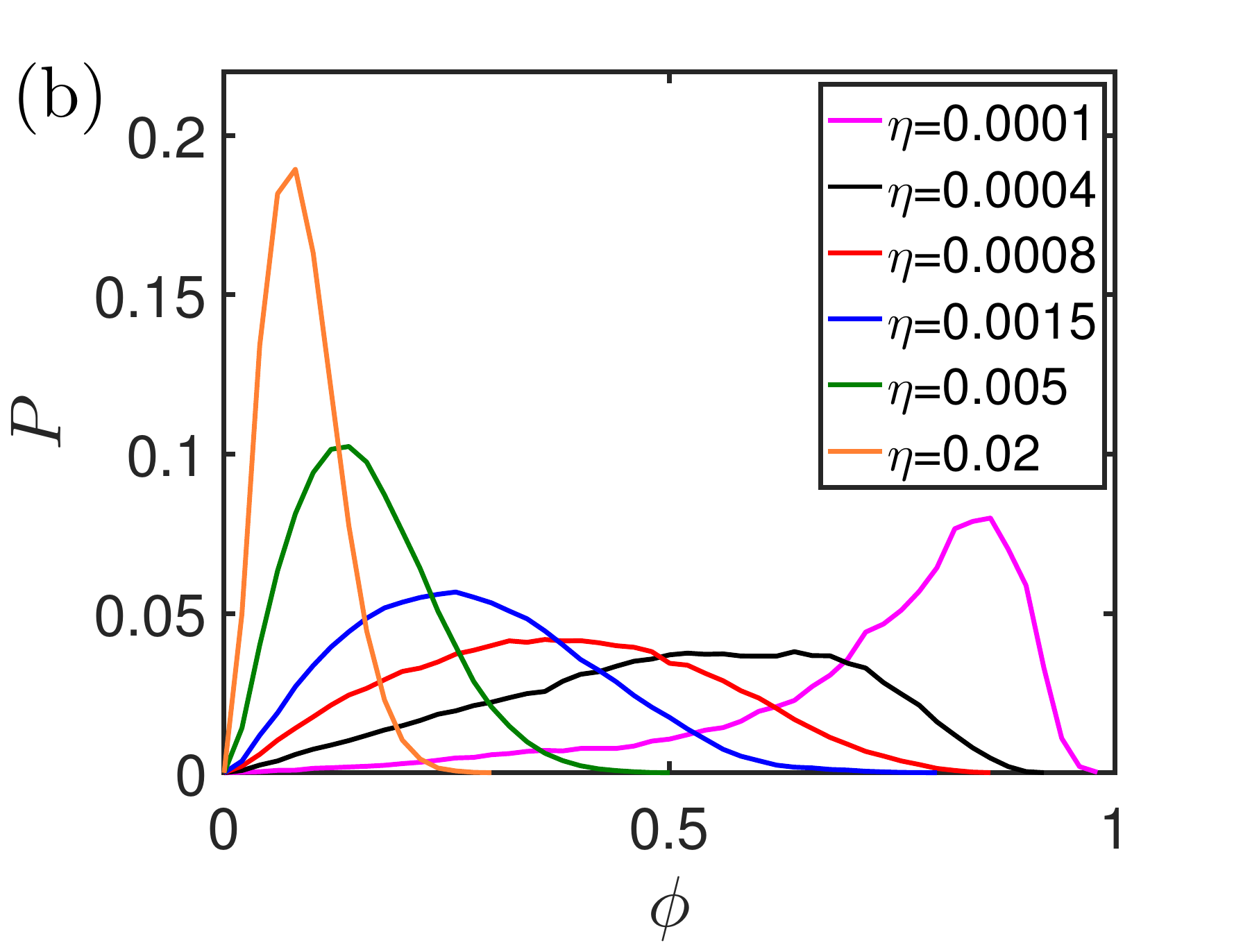}}
\caption{(Color online) Phase transitions of hierarchical swarming model with vectorial noise. 
(a) The average order parameter $\langle\phi\rangle_t$ versus the noise amplitude $\eta$. The chosen values of $\alpha$ are the same as in the case of HVM (Fig. 1) plus $\alpha=1/4$ for comparison.
(b) Probability distribution function of $\phi$ near transition point at $\alpha=0$. 
Parameters: $\rho=2$, $v_0=0.5$, $L=128$.
}
\label{Fig.5}
\end{figure}

The impact of hierarchy on the phase transitions with vectorial noise is shown in Fig.~\ref{Fig.5}a. As seen, an already very strong first-order phase transition is observed without hierarchy ($\alpha=1$). As $\alpha$ decreases, the hierarchy does not further enhance the transition jump as in HVM; the transitions gradually become smoother, continuous-like PTs are seen as $\alpha\rightarrow 0$. To validate, some probability distributions of $\phi$ around the transition point for $\alpha=0$ are shown in Fig.~\ref{Fig.5}b. For all cases, only single-peaked distributions are observed, an unambiguous evidence for continuity. Again, since the critical noise $\eta_c$ is so small that we cannot conclude a strict continuous PT in the thermodynamic limit either.  An intuitive explanation lies in the fact that the presence of hierarchy reduces the local alignment thus the relative intensity of noise becomes stronger, as a result a reduced discontinuity is expected.
Therefore, with the vectorial noise, we only observe that the discontinuity of phase transitions is gradually weakened from egalitarian societies to despotic ones, no even stronger first-order phase transition is detected.

\emph{Conclusions }--- To summarize, we have introduced a simple model for hierarchical swarms and study the impact of hierarchy on the collective movement. As the impact of hierarchy becomes stronger, the society exhibits non-monotonic swarming transitions from weak to strong first-order PTs, and weakened discontinuity again till vanishes, where no order-disorder transition is expected any more in the extremely despotic societies. A two-group model verifies the robustness of these findings. A hierarchical swarm model with vectorial noise is also studied. Microscopically, this is attributed to the altered correlation between the density and orientation fields. 
On the theoretic side, our results points out the importance of the hierarchical attribute when studying swarming transitions; on the experimental side, we expect some case studies with different hierarchies that confirm our conclusion and reveal other complexities induced from hierarchy impact.

We acknowledge Cristi\'an Huepe for many helpful discussions, and financial support from the National Natural Science Foundation of China under Grants 61703257 and 11747309. 

\bibliography{bibfile}
\end{document}